\documentstyle[12pt]{article}
\def\hb{\hbar}   
\def\al{\alpha}

\def\ga{\gamma}
\def\de{\delta}

\def\epc#1#2{\displaystyle E^+_{#1#2}}
\def\epcp#1#2{\displaystyle E^-_{#1#2}}
\def\epo#1#2{\displaystyle F^+_{#1#2}}
\def\epop#1#2{\displaystyle F^-_{#1#2}}

\def\th{\theta}

\def\ka{\kappa}
\def\la{\lambda}

\def\ps{\psi}

\def\Ga{\Gamma}
\def\De{\Delta}

\def\Om{\Omega}

\def\fr#1#2{{{#1} \over {#2}}}
\def\prt{\partial}
\def\pr#1{{#1}^\prime} 
\def\ppr#1{{#1}^{\prime\prime}} 
\def\ap{\al^\prime}

\def\ket#1{|{#1}\rangle}

\def\half{{\textstyle{1\over 2}}}
\def\alf{{1\over 2}}
   
\def\frac#1#2{{\textstyle{{#1}\over {#2}}}}
\def\ni{\noindent}
\def\lsim{\mathrel{\rlap{\lower4pt\hbox{\hskip1pt$\sim$}}
    \raise1pt\hbox{$<$}}}
\def\gsim{\mathrel{\rlap{\lower4pt\hbox{\hskip1pt$\sim$}}
    \raise1pt\hbox{$>$}}}
\def\sqr#1#2{{\vcenter{\vbox{\hrule height.#2pt
         \hbox{\vrule width.#2pt height#1pt \kern#1pt
         \vrule width.#2pt}
         \hrule height.#2pt}}}}

\newcommand{\beq}{\begin{equation}}
\newcommand{\eeq}{\end{equation}}
\newcommand{\bea}{\begin{eqnarray}}
\newcommand{\eea}{\end{eqnarray}}
\newcommand{\rf}[1]{(\ref{#1})}
\newcommand{\eq}[1]{Eq.\ \rf{#1}} 

\renewenvironment{thebibliography}[1]
 { \rm
   \begin{list}{\arabic{enumi}.}
    {\usecounter{enumi} \setlength{\parsep}{0pt}
     \setlength{\itemsep}{3pt} \settowidth{\labelwidth}{#1.}
     \sloppy
    }}{\end{list}}

\begin{document}

\titlepage

\begin{flushright}
{IUHET 322\\}
{November 1995\\}
\end{flushright}
\vglue 1cm
	    
\begin{center}
{
{\bf RADIAL COULOMB AND OSCILLATOR SYSTEMS \\ }
{\bf IN ARBITRARY DIMENSIONS \\}
\vglue 1.0cm
{V. Alan Kosteleck\'y and Neil Russell\\} 
\bigskip
{\it Physics Department\\}
\medskip
{\it Indiana University\\}
\medskip
{\it Bloomington, IN 47405, U.S.A.\\}

\vglue 0.8cm
}
\vglue 0.3cm

\end{center}

{\rightskip=3pc\leftskip=3pc\noindent
A mapping is obtained relating  
analytical radial Coulomb systems 
in any dimension greater than one
to analytical radial oscillators in any dimension.
This mapping, 
involving supersymmetry-based quantum-defect theory,
is possible for dimensions unavailable 
to conventional mappings.
Among the special cases is
an injection from bound states of
the three-dimensional radial Coulomb system 
into a three-dimensional radial isotropic oscillator
where one of the two systems has an analytical quantum defect.
The issue of mapping the continuum states is briefly considered.
 
}

\vskip 1truein
\centerline{\it Accepted for publication in J. Math. Phys.}

\vfill
\newpage

\baselineskip=20pt
{\bf\ni I. Introduction}
\vglue 0.4cm

Various types of correspondence between 
the Kepler-Coulomb and the isotropic-oscillator systems
have been extensively investigated
since the influential work of
Levi-Civita early this century
\cite{lc1906}.
Among the correspondences 
of interest are mappings
that can be constructed between the radial
equations of the quantum systems.
This subject was initiated over 50 years ago
in a paper by Schr\"odinger 
\cite{sch}
addressing the solution
of eigenvalue problems by factorization.
Schr\"odinger discovered a connection
between the radial equation of the 
three-dimensional quantum Coulomb problem
and the radial equation of a $D$-dimensional
quantum harmonic oscillator.
Using a quadratic transformation in the radial coordinate,
he showed that the mapping images all the
states in the three-dimensional discrete Coulomb
spectrum only for oscillators with $D=2$ or $4$.

Schr\"odinger's idea was subsequently 
rediscovered or investigated by a number of authors
\cite{lo,bf,dm,rr,clm}.
An extension relating the radial equations
of the $d$-dimensional Coulomb system and the
$D$-dimensional oscillator for the special case
of even dimensions $d=D$ was given in 
ref.\ \cite{cp}.
A more general mapping for arbitrary $d$ and even $D$
that involves a free parameter was presented in
ref.\ \cite{knt},
along with the corresponding mappings to the 
supersymmetric partners of these systems.
All these correspondences involve oscillators in even
dimensions,
and they incorporate constraints on the allowed range
of angular momenta.
It is possible in general
to map all the states of the $d$-dimensional Coulomb system 
into half the states of a $D$-dimensional oscillator, 
where $d$ is greater than one and $D$ must be even.

Recently,
it has been proposed that some restrictions
on the dimensions or angular momenta
can be removed with the introduction of suitable 
analytical deformations called quantum defects
in one or both systems
\cite{ak}. 
The motivation for this derives from the use 
of supersymmetric quantum mechanics
\cite{ni,wi}
in the context of atomic physics
\cite{kn1},
where supersymmetry-based quantum-defect theory
(SQDT)
\cite{kn}
provides an explicit example with direct physical relevance.

One goal of the present paper 
is to investigate the issue of relaxing
the dimensional constraints
on the radial correspondences via the introduction of  
analytical quantum defects.
The treatment incorporates not only the Coulomb and oscillator
systems but also their supersymmetric partners.
We show that with a suitable choice of defect
it is indeed possible to remove restrictions 
on the mappings.
For instance,
among the examples discussed below is a generalized mapping 
taking any state in the three-dimensional
radial Coulomb problem into a state in an analytically modified 
three-dimensional radial oscillator.
We also briefly consider the continuum states
of the two radial systems.

The focus of this work is the set of radial correspondences
as summarized above.
We do not address here 
the different issue of obtaining surjective mappings
between the full $D$-dimensional oscillator and the
full $d$-dimensional Coulomb systems.
This interesting question has been addressed by a 
number of authors,
originating with 
the parabolic-coordinate transformation of Levi-Civita 
\cite{lc1906}
that relates $D=2$ to $d=2$
and with the mapping 
of Kustaanheimo and Stiefel
\cite{ks}
in their work on celestial mechanics
that relates $D=4$ to $d=3$.
The latter transformation in particular 
has been much investigated
in the quantum context
\cite{cm,bo73,ch80,kin83},
and in recent years extensions 
connecting $D=8$ and $d=5$ have been studied
\cite{dmpst,lk88}. 
While more complete than the purely radial mappings,
all these surjective correspondences are restricted
to a narrow range of dimensions.

The organization of this paper is as follows.
Section II consists primarily of background
on the supersymmetric radial
Coulomb and oscillator systems in arbitrary dimensions
and the known correspondences between them.
It contains key equations needed in the subsequent sections
and provides a perspective useful for our purposes.
In section III,
we introduce SQDT 
for the radial Coulomb and oscillator systems 
in arbitrary dimensions,
and we define mappings relating these systems to 
the supersymmetric sectors of the associated zero-defect cases.
The general correspondence between different SQDT 
for the Coulomb and oscillator systems
in arbitrary dimensions is established in section IV.
This permits,
for instance,
the entire set of Coulomb radial states
to be injected into a subset of the oscillator radial states
for any dimensions,
including odd oscillator dimensions.
Section V provides a short discussion of some results
arising for the continuum Coulomb states.
We summarize in section VI.

To distinguish comparable quantities in the two systems,
we adopt the convention that 
lower-case letters are used for Coulomb-system variables
while 
upper-case letters are used for oscillator-system variables.
An exception is made in denoting energies,
for which the symbol $E$ with various sub- and superscripts
is used in the Coulomb system 
while $F$ is used in the oscillator system.

\vglue 0.6cm
{\bf\ni II. Preliminaries}
\vglue 0.4cm

This section establishes our conventions and
presents some preliminary material and results.
Section IIA begins 
with definitions and solutions for the radial 
Coulomb problem in arbitrary dimensions,
while section IIB
similarly treats the harmonic-oscillator radial problem.
A one-parameter mapping between these systems is
presented in section IIC.
Key equations for supersymmetric quantum mechanics 
are given in section IID.
The supersymmetric counterpart
of the results in section IIA is discussed in section IIE,
while that of sections IIB and IIC is covered in section IIF.

\vglue 0.6cm
{\bf\ni IIA. Coulomb Bound States in $d$ Dimensions}
\vglue 0.4cm

The quantum Kepler-Coulomb system in $d$ dimensions
is governed by the hamiltonian
\beq
h = - \fr{\hb^2}{2 \mu} \nabla^2 - \fr \ka r
\quad ,
\label{a2}
\eeq
where $\mu$ is the reduced mass,
$\ka$ is the force constant,
and $r$ is the usual radial variable. 
To avoid normalization issues,
we assume $d>1$.
The associated radial equation is obtained 
from the Schr\"odinger equation $h \ps = E \ps$
by separating the wave function $\ps$ 
in generalized polar coordinates,
$\ps(r,\th_1,\ldots,\th_{d-1}) 
= s(r) \th(\th_1,\ldots,\th_{d-1})$.
Normalizable solutions to the ensuing 
radial equation are found for discrete eigenenergies
given by
\beq
E_{n, \ga} = \fr {- E_0}{4(n + \ga)^2}
\quad ,
\label{a6} 
\eeq
where 
$E_0 = 2 \mu \ka^2 / \hb^2$,
$\ga = \half (d-3)$,
and $n$ is the principal quantum number taking values
$n=l+1$, $l+2$, $\ldots $,
with $l$
the angular-momentum quantum number
arising from the separation of variables.
For $d\ge 3$, 
$l = 0$, $1$, $2$, $\ldots$ as usual.
When $d=2$,
the angular-momentum quantum number
takes the values $0$, $\pm 1$, $\pm 2$, $\ldots$.
In this case,
we define the symbol $l$ to 
represent the modulus of the angular momentum.
Note that $E_0$ is the magnitude 
of the ground-state energy in the 
lowest dimension considered, $d=2$.

To simplify the equations that follow,
we introduce a dimensionless radial variable
$y = r/r_0$,
where $r_0 = \hb^2 / 2 \ka \mu$.
It is also convenient for later considerations
involving supersymmetry to work with 
a scaled radial function
$w(y) = y^{\ga+1} s(r\equiv r_0y)$,
which effectively removes the first-order derivative
appearing in the radial equation for $s$.
The radial equation becomes
\beq
\left\{ - \fr{d^2}{d y^2}  + \fr{(l+\ga)(l+\ga+1)}{y^2}
- \fr 1 y - \fr E {E_0} \right\} w_{d,n,l}(y) = 0
\quad .
\label{a5}
\eeq
The eigensolutions involve Sonine-Laguerre polynomials
and are given by
\beq
w_{d,n,l}(y) = c_{dnl} \, y^{l+\ga+1} 
\exp (\frac{- y}{2(n+\ga)}) \, 
 L^{(2l+2\ga+1)}_{n-l-1}(\fr y {n+\ga})
\quad ,
\label{a7}
\eeq
with normalization
\beq
c_{dnl} = \left[ 
\fr{\Ga(n-l)}{2 r_0^d (n+\ga)^{2l+d+1} \Ga(n+l+d-2)} 
\right]^\half
\quad .
\label{a8}
\eeq

\vglue 0.6cm
{\bf\ni IIB. Oscillator Bound States in $D$ Dimensions}
\vglue 0.4cm

The quantum hamiltonian for 
the isotropic harmonic oscillator in $D$ dimensions,
$D\ge 1$,
is 
\beq
H = - \fr{\hb^2}{2 M} \nabla^2 + \half M \Om^2 R^2
\quad ,
\label{b2}
\eeq
where $M$ is the oscillator mass,
$\Om$ is the frequency,
and $R$ is the usual radial variable. 
Separating variables in generalized polar coordinates
as before produces a radial equation 
that has normalizable solutions $S(R)$
for energy eigenvalues given by
\beq
F_{N, \Ga} = F_0 (2N + 2\Ga + 3)   
\quad ,
\label{b4}
\eeq
where 
$\Ga = \half(D-3)$, 
$F_0 = \half \hbar \Om$
is the ground-state energy for the lowest dimension $D=1$, 
and $N$ is the principal quantum number taking values
$N=L$, $L+2$, $L+4$, $\ldots$,
with $L$ the quantized angular momentum
arising from the separation of variables.
For $D\ge 3$, 
$L = 0$, $1$, $2$, $\ldots$ as usual.
For $D=2$, 
the angular momentum ranges over 
$0$, $\pm 1$, $\pm 2$, $\ldots$,
and we define $L$ to be its modulus.
For $D=1$,
the only possibilities are $L=0$ and $1$.
The corresponding angular variable has two discrete values,
distinguishing the two orientations of the radial vector.
These two cases represent distinct \it radial \rm systems
for the one-dimensional oscillator,
as opposed to the full one-dimensional oscillator
with configuration space including both positive and negative 
coordinate values.
For convenience in what follows,
we define the parity of the radial wave functions
as even if $L=0$ and odd if $L=1$.

Defining the dimensionless variable
$Y=R/R_0$ with $R_0 = (\hb / M \Om)^\half$
and introducing for later convenience the scaled
radial function $W(Y) = Y^{\Ga+1}S(R \equiv R_0 Y)$,
the radial equation becomes
\beq
\left\{ - \fr{d^2}{dY^2} + \fr{(L+\Ga)(L+\Ga+1)}{Y^2} + 
Y^2 - \fr{F}{F_0} \right\} W(Y) = 0
\quad .
\label{b3}
\eeq
The eigenfunctions are
\beq
W_{D,N,L}(Y) = C_{DNL} \, Y^{L+\Ga+1} \exp (- \half Y^2)
L^{(L+\Ga+\half)}_{\fr N 2 - \fr L 2} (Y^2)
\quad ,
\label{b5}
\eeq
with normalization
\beq
C_{DNL} = \left[\fr{2 \Ga(\fr N 2 - \fr L 2 + 1)}
{R_0^D \Ga(\fr N 2 + \fr L 2 + \fr D 2)}\right]^\half
\quad .
\label{b6}
\eeq
With our definitions for $L$ above,
these expressions hold for all integral $D\ge 1$.
Note that the $D=1$ normalization coefficients differ from
the canonical ones by a factor of $\sqrt 2$ because 
the above construction produces a normalization 
on the half line only.

\vglue 0.6cm
{\bf\ni IIC. Mappings between 
the Coulomb and Oscillator Problems}
\vglue 0.4cm

The wave function \rf{a7} can be mapped to the
wave function \rf{b5} through the quadratic transformation 
\beq
Y^2 = \fr y {(n+\ga)} 
\quad .
\label{c0}
\eeq
This correspondence also interconnects 
the differential equations
\rf{a5} and \rf{b3}.
The explicit relation between eigensolutions is
\beq
W_{D,N,L}(Y) =  K_{d n \la} Y^{- \alf}
     \ w_{d,n,l}\left((n+\ga)Y^2\right) 
\quad ,
\label{c1} 
\eeq
where the quantity $K_{d n \la}$ maintains the
normalization of the wave functions and is given by
\beq
K_{d,n,\la} =  \fr{(2n+d-3) ~r_0^{d/2}}
{R_0^{d-1-\la}} 
\label{c2}
\quad .
\eeq
The quantity $\la$ provides an extra degree of freedom
in the mapping.

The ensuing relationships among the dimensionalities
and the quantum numbers of the two systems are
\cite{knt}
\bea
D &=& 2d - 2 - 2\la \label{c3a}
\quad , \nonumber\\
N &=& 2n - 2 + \la \label{c3}
\quad , \nonumber\\
L &=& 2l + \la  \label{c3b} 
\quad .
\label{rels}
\eea
The last of these equations constrains $\la$ to be integral.
It then follows from the first equation 
that this mapping has image 
only in the oscillators of even dimension $D$.

For given angular momenta $l$ and $L$,
the relation \rf{c3} between the principal quantum numbers
ensures that the stack $n \geq l+1$ of Coulomb states 
is in one-to-one correspondence with 
the stack $N \geq L$ of oscillator states,
with ground states coinciding.
This relation between $N$ and $n$ 
determines a condition relating 
the energies $E$ and $F$ of the two systems:
\beq
\fr {F_{N, \Ga}}{F_0} = 2 \, \sqrt{\fr{E_0}{-E_{n, \ga}}}
\quad .
\label{c4}
\eeq
The factor of two can be viewed as 
originating from the scaling of $N$ relative to $n$
in the second equation in \rf{rels}.
Absorbing it in the definitions of $E_0$ or $F_0$
would change equations \rf{a5} or \rf{b3}.

Condition \rf{c3b} shows that successive 
Coulomb angular momenta $l$
map to every second oscillator angular momentum $L$. 
The \it entire \rm set of radial states $\ket{n, \, l}$ 
of the $d$-dimensional Coulomb system 
can be mapped into a subset of the states 
$\ket{N, \, L}$ of the $D$-dimensional oscillator 
provided $D$ satisfies
$2 \leq D \leq 2d-2$.
For even or odd $\la$,
the mapping is then an isomorphism
to even or odd $L$,
respectively.
For given $d$,
the allowed values of the pair $(D,\la)$
characterizing this mapping are distinct: 
$(2,d-2)$,
$(4,d-3)$,
$\ldots$, 
$(2d-4,1)$, 
$(2d-2,0)$. 
We recover in this way Schr\"odinger's result
that all states of the three-dimensional Coulomb system 
can be mapped only into oscillators 
of dimension two or four.
Note that
if instead $l$ is taken to be fixed,
so that only a subset of states is imaged,
then the allowed range of $D$ is
\cite{bf}
$2 \leq D \leq 2d-2+4l$.

\vglue 0.6cm
{\bf\ni IID. Supersymmetric Quantum Mechanics}
\vglue 0.4cm

For the purposes of the present paper,
only a few of the basic results of 
supersymmetric quantum mechanics
are needed.
We restrict our attention
to systems with
a quantum-mechanical hamiltonian $H_{S}$ 
and two supersymmetry charges $Q$ and $Q^{\dag}$,
obeying the defining relations
of the superalgebra sqm(2):
\beq
\{Q,Q^{\dag} \} = H_{S}
\quad , \qquad
[Q, H_{S}] = [Q^{\dag}, H_{S}] = 0
\quad .
\label{d3}
\eeq

The representation of this algebra relevant here is 
two dimensional and may be parametrized as
\cite{ni,wi}
\beq
Q = 
\left(\begin{array}{cc} 
    0 & 0 \\ A & 0 
\end{array}\right) 
\quad , \ \ \
Q^{\dag} = 
\left(\begin{array}{cc} 
    0 & A^{\dag} \\ 0 & 0 
\end{array}\right)
\quad , \ \ \
H_{S} = 
\left(\begin{array}{cc} 
      H^+ & 0 \\ 0 & H^-  
\end{array}\right) 
\quad . 
\label{d4}
\eeq
There are two component hamiltonians in this system
and two associated Hilbert spaces.
If the bosonic hamiltonian $H^+$ 
acts on wave functions $\ps^+$,
while the fermionic hamiltonian $H^-$ acts on $\ps^-$,
then the corresponding Schr\"odinger equations can be written
\beq
H^\pm \ps^\pm 
= \left[ - \fr{d^2}{dy^2} + V^\pm (y) \right] \ps^\pm
= E^\pm \ps^\pm
\quad ,
\label{d5}
\eeq
where $A$ is the operator $A = -i \prt_y - i U^\prime$ and 
where the supersymmetric partner potentials are defined by
$V^\pm (y) = U^{\prime 2}\mp U^{\prime\prime}$,
with $U^\prime = \prt_y U(y)$ a specified function called
the superpotential.

The ground state of a supersymmetric system 
lies in the bosonic sector and has zero energy. 
Every state in the bosonic sector other than the ground state 
is degenerate with a distinct state in the fermionic sector,
and the operators $Q$, $Q^{\dag}$ 
map between these paired states.

\vglue 0.6cm
{\bf\ni IIE. Supersymmetric Coulomb System}
\vglue 0.4cm

To construct the supersymmetric Coulomb system,
the bosonic-sector combination $H^+ - E^+$ from Eq.\ \rf{d5}
is identified with the radial equation \rf{a5}.
In the latter,
a suitable constant must be added to the energy eigenvalues 
and incorporated in the potential
to ensure that the ground-state energy is zero.
Thus, 
the eigenvalues are 
\beq
\epc nl = \fr 1 {4(l+1+\ga)^2} - \fr 1 {4(n+\ga)^2}
\quad ,
\label{e1}
\eeq
and the bosonic potential function is 
\beq
v^+(y) = 
\fr{(l+\ga)(l+\ga+1)}{y^2} - \fr 1 y + \fr1 {4(l+1+\ga)^2}
\quad .
\label{ee1}
\eeq
The last term in Eq.\ \rf{ee1} 
is the energy shift 
ensuring a zero ground-state energy.
Since it must be constant,
$l$ must be fixed to define a supersymmetric partner.

The superpotential is specified by the function
\cite{kn1}
\beq
u(y) = \fr y {2(l+\ga+1)} - (l+\ga+1)\ln y
\quad .
\label{e2}
\eeq
The fermionic hamiltonian and hence the associated
fermionic radial equation can then be calculated as 
\beq
\left\{ - \fr{d^2}{d y^2}  
+ \fr{(\pr l +\ga)(\pr l+\ga+1)}{y^2}
- \fr 1 y + \fr 1 {4(\pr l +\ga)^2} \right\} w^-(y) 
= \epcp {\pr n}{\pr l} \ w^-(y)
\quad ,
\label{ee2}
\eeq
where $\pr l =l+1$ and 
$n^\prime$ takes on all values of $n$ 
except the lowest one, $l+1$.
The fermionic wave functions $w^-(y)$ have 
the same functional form as the bosonic wave functions
$w^+(y) \equiv w(y)$ given in Eq.\ \rf{a7},
but with $n$ and $l$ replaced by $\pr n$ and $\pr l \,$:
\beq
w^-_{d, \pr n, \pr l} (y) =  w_{d, \pr n, \pr l} (y)
\quad .
\label{ee3a}
\eeq
The two sets of eigenvalues 
are degenerate for $\pr n = n$: 
$\epcp {\pr n=n,}{\pr l} = \epc n l \,$.

With fixed $l$,
the bosonic stack of eigenstates in order of increasing energy
consists of the series of kets
$\ket{n=l+1,\ l}$, $\ket{n=l+2,\ l}$, $\ldots$, 
with lowest energy zero. 
The associated fermionic stack 
has angular momentum greater by one unit,
and starts with lowest energy corresponding 
to the second state of the bosonic sector:
$\ket{\pr n =l+2,\ l+1}$, 
$\ket{\pr n =l+3,\ l+1}$, 
$\ldots$.
The use of $\pr n$ here 
is consistent with spectroscopic notation.
For example, 
when the s orbitals of lithium
are interpreted as the supersymmetric partner 
of the hydrogen atom,
$\pr n=2$ corresponds to the ground state,
as expected.

A useful one-to-one correspondence between these two stacks 
identifies the lowest states with each other,
and successively higher states of the bosonic sector 
with successively higher states of the fermionic sector.
It is defined by the following replacements 
in the bosonic wave function:
\bea
n & \longmapsto & \pr n = n + 1  
\quad , \nonumber \\
l & \longmapsto & \pr l = l + 1  
\quad .
\label{e3}
\eea
This stack correspondence relates eigenstates with
different eigenvalues.
Along with similar stack correspondences defined below,
it plays a useful role in the analyses to follow.

\vglue 0.6cm
{\bf\ni IIF. Supersymmetric Oscillator 
and Composition Mapping}
\vglue 0.4cm

The bosonic component of the supersymmetric oscillator
can be obtained from the radial equation \rf{b3}
under a suitable energy shift.
The eigenvalues are 
\beq
\epo N L = 2 (N-L)
\quad .
\label{f1}
\eeq
The superpotential is specified via the function
\beq
U(Y) = \half Y^2 - (L + \Ga +1) \ln Y 
\quad ,
\label{f2}
\eeq
which generates the fermionic equation
\beq
\left\{ -\fr{d^2}{dY^2} + \fr{(\pr L+\Ga)(\pr L+\Ga+1)}{Y^2} + 
Y^2 - (2\pr L+2\Ga-1) \right\} W^- (Y) 
= \epop {\pr N}{\pr L} W^-(Y)
\label{ff2}
\quad ,
\eeq
where $\pr L$ is defined by $\pr L = L+1$.
The principal quantum number $\pr N$ 
takes the values 
$\pr N = \pr L$, $\pr L +2$, $\pr L +4$, $\ldots $.
The fermionic wave functions $W^-(y)$ have 
the same functional form as the bosonic 
wave functions $W^+(Y)\equiv W(Y)$ of 
Eq.\ \rf{b5},
with $N, \ L$ replaced by $\pr N, \ \pr L$,
respectively:
\beq
W ^-_{D, \pr N, \pr L}(Y) =  W _{D, \pr N, \pr L}(Y) 
\label{ff3a}
\quad .
\eeq
The fermionic energies 
$\epop {\pr N}{\pr L} = 2 (\pr N - \pr L + 2)$
are degenerate with the bosonic energies 
for $\pr N + 1 = N$,
$\epop {\pr N}{\pr L} = \epo {N=\pr N + 1,}{L}$.

In order of increasing energy, 
the bosonic stack with fixed $L$ consists of the kets
$\ket{N=L, \ L}$, $\ket{N=L+2,\ L}$, $\ket{N=L+4, \ L}$, 
$\ldots$, 
with lowest energy zero. 
The associated fermionic stack 
has angular momentum one unit greater 
and contains the kets 
$\ket{\pr N=\pr L =L+1,\ L+1}$, $\ket{\pr N =L+3,\ L+1}$, 
$\ldots$.

The $D=1$ case is unusual and warrants special attention.
As mentioned above, 
the `angular momentum' $L$ 
for the one-dimensional oscillator
takes the values zero and one,
corresponding to even and odd parity.
The system resembles a single stack, 
but is composed of two interlocking substacks.
As a result,
the spacing between neighboring eigenvalues 
is half its value in higher dimensions.
Also,
in constructing the supersymmetric partner,
the energy shift is $L$ dependent.
Consequently, 
distinct shifts appear for each substack.
The formalism thus establishes 
\it two \rm independent supersymmetries, 
each of which respects the parity
and only one of which may be considered at a time.
These supersymmetries for the 
one-dimensional radial oscillator
differ from the usual one 
for the full one-dimensional oscillator,
where a single energy shift is effected 
and states of opposite parity are degenerate
under the supersymmetry.

In later sections,
for reasons that emerge from the construction of
the generalized mapping,
it is more natural to focus on 
the supersymmetric partner of the fermionic oscillator
rather than the fermionic oscillator itself.
This system,
which we call `second fermionic,'
has wave functions 
$W^=_{D\ppr N\ppr L}$
given by
\beq
W^=_{D\ppr N\ppr L}(Y) = W_{D,\ppr N,\ppr L}(Y)
\quad .
\label{f5}
\eeq
Here,
$\ppr L$ is defined by $\ppr L = \pr L +1$,
and 
$\ppr N$ takes values $\ppr L$, $\ppr L+2$,$\ppr L+4$,
$\ldots$.
The differential equation for this system
has the same functional form as
the fermionic equation \rf{ff2},
except for the replacement of $N$ and $L$
with $\ppr N$ and $\ppr L$,
respectively.

The oscillator bosonic sector may be put into 
one-to-one correspondence with the second-fermionic sector 
by making the following replacements 
in the bosonic wave function:
\bea
N & \longmapsto & \ppr N = N+2 
\quad ,
\nonumber \\
L & \longmapsto & \ppr L = L+2
\quad .
\label{f6}
\eea

By composition of this mapping and the ones given in sections 
IIC and IIE,  
a correspondence may be established between the 
fermionic sector of the Coulomb system and
the second-fermionic sector of the oscillator.
It is given by
\bea
W^=_{D,\ppr N, \ppr L}(Y) 
&=& K_{d,\pr n , \la} Y^{- \alf} 
\ w^-_{d \pr n \pr l }
          \left((\pr n  +\ga)Y^2\right) \quad ,  \\ 
Y^2 & = & \fr y {(\pr n  + \ga)}  
\quad , \\
\ppr N & = & 2 \pr n  - 2 + \la \quad , \\
\ppr L & = & 2 \pr l  + \la 
\quad .
\label{f7}
\eea
The dimensions are still related as in 
Eq.\ \rf{c3a}. 
See Figure 1.
We emphasize that this commutative diagram 
involves mappings different from those
presented in ref.\ \cite{knt},
where the second-fermionic sector is not considered.

\vglue 0.6cm
{\bf\ni III. Generalized Supersymmetry-Based 
Quantum-Defect Theory}
\vglue 0.4cm

In this section,
we introduce analytical SQDT 
for the Coulomb and oscillator systems
in arbitrary dimensions.
{}From the present perspective,
the goal is to obtain effective radial equations
that offer sufficient flexibility to obviate the 
dimension and angular-momentum constraints of the usual mappings,
while maintaining eigensolutions with 
analytical structure comparable in simplicity
to those of the Coulomb and oscillator systems.

The existence of suitable deformations
of the Coulomb and oscillator systems
satisfying these criteria is by no means apparent 
{\it a priori}.
In what follows,
we take as a guide the SQDT that is known 
to provide a useful analytical description
of the valence structure of physical atoms
in terms of an effective one-particle radial equation
\cite{kn}.
This model determines an effective radial potential
modifying the three-dimensional radial Coulomb equation.
It generates solutions with physical eigenvalues
given by the Rydberg expression 
\beq
E_{n^*}= -E_0/4 n^{*2}
\quad .
\label{ryd}
\eeq
Here, 
$n^*$ is the principal quantum number
modified by subtracting the quantum defect $\de$,
which in general depends on 
the angular momentum
and the principal quantum number.
In section IIIA,
we generalize this model to the $d$-dimensional situation.
For simplicity,
we take $\de$ and its generalization 
in arbitrary dimensions 
to be independent of the principal quantum number.
This approximation is excellent in,
for example,
real alkali-metal atoms
\cite{hk}. 

Similar ideas can be implemented 
for the radial equation of the $D$-dimensional oscillator.
The resulting oscillator SQDT are
presented in section IIIB.
A possible physical application of these oscillator models is 
to the description of a valence particle
in geonium atoms formed by a group 
of charged particles bound in a Penning trap
\cite{ak}.

\vglue 0.6cm
{\bf\ni IIIA. Generalized SQDT for the Coulomb System}
\vglue 0.4cm

Given the $d$-dimensional
Coulomb radial equation \rf{a5}
with fixed angular momentum $l$,
we seek to implement two modifications 
via an effective potential $v_{\rm eff}(y)$
added to the left-hand side.
The first desired modification is a shift in dimension,
from $d$ to $d^* = d+j$, 
where $j$ is an integer
that in principle could depend on $l$.
We require $d^* > 1$, 
so $j$ must satisfy $j > 1-d $.
The second desired modification is 
a shift in energy eigenvalues
from $E_{n,\ga}$ in Eq.\ \rf{a6}
to the $d^*$-dimensional extension of the Rydberg series
(see \eq{CouQDEn} below).
We want both these changes to be implemented
while maintaining analytical eigenfunctions
with form similar to those in Eq.\ \rf{a7}.

Remarkably,
these goals can be accomplished 
with a relatively simple effective potential,
given by
\beq
v_{\rm eff} (y) = 
\fr{(n+\ga)^2-(n^* +\ga^*)^2}{4(n +\ga)^2 (n^* +\ga^*)^2}
+ \fr{(l^* +\ga^*)(l^* +\ga^*+1)-(l+\ga)(l+\ga+1)}{y^2}
\quad .
\label{g1}
\eeq
Here,
the quantity $\ga^*$ is defined by 
$\ga^* = (d^*-3)/2$.
The quantities $n^*$ and $l^*$  are defined as 
\bea
n^* &\equiv& n_s - \de \ = \ n+i-\de
\quad , \qquad \label{g1a0} \\
l^* &=& l+i-\de
\quad ,
\label{g1a}
\eea
where $\de$ is the quantum defect determining the
energy shifts for the generalized Rydberg series
and where $i=i(l)$ is an integral-valued function of
the angular momentum.
In the supersymmetric interpretation
for the valence electron of physical atoms,
$i(l)$ is the number of 
filled lower levels with angular momentum $l$.
The introduction of $n_s$ is motivated 
by the three-dimensional case,
where it is equal to the principal quantum number
and takes conventional values 
in the standard spectroscopic notation.
It satisfies $n_s=n+i$,
where $n$ takes the usual values
characteristic of the exact Coulomb system.
As an example, 
the $s$ states of the supersymmetric sodium atom 
in three dimensions have $i(0)=2$, 
giving $n_s=3$, $n_s=4$, $n_s=5$ 
for the first three levels
\cite{kn1}.
The corresponding values of $n$  
are $n=1$, $n=2$, $n=3$.
For the supersymmetric partner
of the exact Coulomb system,
$i=1$ and so $n_s = n+1 = \pr n$,
consistent with our previous notation 
for the supersymmetric case.

The first term 
of Eq.\ \rf{g1} has the effect 
of shifting the energy levels,
while the second term performs a corresponding shift in
the angular-momentum barrier.
The combined effect of both terms incorporates 
the desired dimensional shift.
With a nonzero quantum defect $\de(l)$,
the effective potential $v_{\rm eff}(y)$ plays the role of
a supersymmetry-breaking potential.
The resulting radial equation has analytical solutions
given in terms of the usual Coulomb solutions 
$w_{d,n,l}(y)$ by $w_{d^*,n^*,l^*}(y)$.
These solutions exist for $n \ge l+1$, 
or $n_s \ge l+i+1$. 
Requiring the existence and
orthonormalizability of the wave functions 
restricts $\de - i$ according to 
\bea
\de-i < l+\ga+1 + \half j
\quad .
\label{cRange}
\eea

It is convenient to define a dimensionless quantity 
$a(l)$ by 
\beq
a(l) = i - \de + \half j \quad .
\label{g1b}
\eeq
The eigenvalues of the differential equation 
can then be expressed as
\beq
\fr{E_{n^*,\ga^*}}{E_0} \ = \ \fr {-1}{4(n^*+\ga^*)^2}
                          \ = \ \fr {-1}{4(n+\ga+a)^2}
\quad .
\label{CouQDEn}
\eeq
In this equation,
we have chosen the eigenenergies
so that the limiting case with $d=3$ and $i=j=0$ reproduces 
the Rydberg series \rf{ryd}.
For $i$, $\de$, and $j$ chosen so that $a=0$,
we obtain the bosonic equation of the Coulomb system
discussed in section IIE,
up to an energy shift.
If $a=1$, the fermionic sector
of the Coulomb problem is generated instead.
Moreover,
the supersymmetric partner of the fermionic sector 
is generated by setting $a=2$, 
and each successive iteration of the supersymmetry 
increments $a$ by one unit.

A useful stack correspondence can be established
between the spectrum 
of the bosonic sector of the supersymmetric
Coulomb system and the SQDT Coulomb spectrum.
The map is given by making 
the following replacements in $w^+_{d,n,l}$:
\bea
d & \longmapsto &  d^* = d + j 
\quad , \nonumber \\
n & \longmapsto &  n^* = n + i - \de = n_s - \de 
\quad , \nonumber \\
l & \longmapsto &  l^* = l + i - \de  
\label{g3}
\quad .
\eea

{\bf\ni IIIB. Generalized SQDT for the Oscillator System}
\vglue 0.4cm

The techniques of section IIIA
can also be applied to the radial oscillator system
in $D$ dimensions.
For fixed angular momentum $L$,
we can obtain an effective potential $V_{\rm eff}(Y)$
to be added to Eq.\ \rf{b3} 
that maintains analytical eigenfunctions
while inducing an integral shift 
to a new dimension $D^* \equiv D+J \geq 1$
and simultaneously modifying the oscillator energy eigenvalues
via a shift to a new principal quantum number $N^*$.
We refer to the resulting theory as the oscillator SQDT.

The appropriate choice of effective potential is
\beq
V_{\rm eff} (Y) = 
2(N - N^* + \Ga - \Ga^*)
+ \fr{(L^* +\Ga^*)(L^* +\Ga^*+1)-(L+\Ga)(L+\Ga+1)}{Y^2} 
\quad ,
\label{h1}
\eeq
where
$\Ga^* = (D^*-3)/2$
and the shifted quantum numbers are given by 
\bea
N^* &\equiv & N_s - I - \De \equiv N+I-\De       
\quad ,\label{h1a} \\
L^* &=& L + I - \De
\quad .
\label{h2}
\eea
Here,
$I=I(L)$ is an integral-valued function,
analogous to $i(l)$ in the Coulomb case,
that can be interpreted as the number of
inaccessible lower levels.
The quantity $\De (N,L)$ is the oscillator equivalent
of the Rydberg quantum defect $\de (n,l)$,
modifying the radial-repulsion term in the differential equation.
For simplicity in what follows,
we take $\De$ to depend only on $L$,
thereby paralleling the case of alkali-metal atoms
for which $\de$ depends only on $l$.
We have also defined a quantity 
$N_s$ playing the role of the principal quantum number 
in the spectroscopic notation, 
given by $N_s= N+2I$. 
If the dimension is unmodified and $\De=0$, 
the choice $I=1$ yields the fermionic sector 
of the supersymmetry discussed in section IIF.
In this limit $\pr N = N^* = N+1 \neq N_s$,
which differs from the supersymmetric limit of 
the Coulomb SQDT where $\pr n = n^* = n_s$.
With our definitions,
degenerate levels in the bosonic and fermionic sectors
have values of $N$ differing by $2I$ units,
but have the same value of $N_s$.

The first term in Eq.\ \rf{h1}
implements the eigenenergy shift
to the oscillator analogue of the Rydberg series,
while the second term is
the corresponding anharmonic modification
to the potential that maintains analytical eigensolutions.
The eigenfunctions solving the resulting effective
radial equation are given in terms of the
oscillator wave functions $W_{DN,L,}(Y)$
of Eq. \rf{b5} by
$W_{D^*,N^*,L^*}(Y)$.
The existence of these solutions requires that 
the principal quantum number takes the values
$N_s=L+2I$, $L+2I+2$, $L+2I+4$, $\ldots$, or
$N=L$, $L+2$, $L+4$, $\ldots$.
Requiring orthonormalizability of the wave functions
restricts the range of $\De - I$ to
\beq
\De - I < L + \Ga + \frac 3 2 + \half J
\quad . \label{oRange} 
\eeq

We can again introduce 
a useful dimensionless quantity $A(L)$ by
\beq
A(L) = I -\De + \half J
\quad .
\label{h2c}
\eeq
The eigenvalues of the differential equation 
can be expressed as
\beq
\fr{F_{N^*,\Ga^*}}{F_0} \ = \  2N^* + 2\Ga^* +2A + 3
                         \ = \ 2N +2\Ga +4A +3
\quad .
\label{OscQDEn}
\eeq
We have chosen the ground-state eigenenergy
in analogy with the Coulomb case \rf{CouQDEn}.
The extra factor of $2A$ appears to ensure that the 
bosonic and fermionic spectra of the limiting
supersymmetric case with $\De = 0$, $J=0$
have the characteristic degenerate pairing.
If $I$, $\De$, and $J$ are selected so that 
$A =0$, 
then this SQDT system reduces to 
the bosonic oscillator discussed in section IIF,
up to an energy shift.
If $A =1$, 
it reduces instead to the fermionic partner.
If $A=2$,
the second-fermionic sector of the supersymmetric oscillator 
is produced. 
Each further iteration of supersymmetry
produces an additional unit increment of $A$.
Note that for fixed $L$ 
the spacing between successive eigenvalues 
is always four units,
regardless of the value of $A$.

A correspondence
can be established between the 
oscillator SQDT and the 
bosonic sector of the supersymmetric oscillator. 
The images of the wave functions
are obtained by making the replacements:
\bea
D & \longmapsto &  D^* = D + J
\quad , \nonumber \\
N & \longmapsto &  N^*  = N + I - \De 
\quad , \nonumber \\
L & \longmapsto &  L^* = L + I - \De 
\quad .
\label{h6}
\eea

\vglue 0.6cm
{\bf\ni IV. Mappings between Bound States 
of the Coulomb and Oscillator SQDT}
\vglue 0.4cm

Composition of the mappings in sections
IIC, IIIA and IIIB
allows us to establish a correspondence
between the $d^*$-dimensional Coulomb SQDT and 
the $D^*$-dimensional oscillator SQDT. 
This mapping is described in section IVA.
One of its striking features is 
that the odd-dimensional oscillator can be imaged.
In section IVB,
we illustrate the mapping with
examples involving the three-dimensional
Coulomb and oscillator systems.

\vglue 0.6cm
{\bf\ni IVA. The General Case}
\vglue 0.4cm

The general mapping is given by
\bea
 W_{D^*,N^*,L^*}(Y) &=&
     K_{d^*,n^*,\la-\alf J + j} 
     Y^{-\alf} 
     w_{d^*,n^*,l^*} \left((n^*+\ga^*)Y^2\right)  
           \quad , \label{jjb} \\
Y^2 & =&  y /(n^* +\ga^*)\quad ,\\
D^* &=& 2 d^* - 2 - 2 \la +  J - 2j \quad , \label{jjd} \\
N^* &=& 2 n^* -2+\la-\half J+j   \quad , \label{j1}  \\
L^* &=& 2 l^*+\la-\half J+j\label{jjc} \quad ,\\ 
A &=& 2 \, a \quad . \label{jje}
\eea
This mapping,
like the Coulomb-oscillator case discussed in section IIC,
is based on a quadratic relationship 
between the radial variables of the two systems.
The constant $K$ can be chosen 
to preserve the normalization of the wave functions,
in which case it has the functional form given in \eq{c2}.
Note that \eq{j1} is equivalent to a 
generalization of \eq{c4},
given by
\beq
\fr {F_{N^*, \Ga^*}}{F_0} = 
2 \, \sqrt{\fr{E_0}{-E_{n^*, \ga^*}}} + 4a
\quad .
\label{}
\eeq
Note also that
the allowed ranges of the quantum defects given in 
\eq{cRange} and \eq{oRange}
are compatible with \eq{jje},
which guarantees that the image of 
any orthonormalizable Coulomb radial system 
is an orthonormalizable oscillator.

To gain insight about the flexibility of this mapping,
consider the choice $j=0$.
Then,
$D^*$ lies in the range $1+J \leq D^* \leq 2d-2+J$
with allowed values separated by two units. 
Since we require $J \geq 1-D$,
any $D^* \geq 1$ is possible.
Moreover,
$d$ may take any value greater than one.
The above general mapping therefore 
relates any Coulomb dimension $d > 1$
to any oscillator dimension $D^* \geq 1$.
In particular,
\eq{jjd} shows that $D^*$ is odd if 
$J$ is chosen to be an odd integer. 
This is in striking contrast to the 
usual restriction of $D$ to even values only,
as given by \eq{c3a}.

The Coulomb-oscillator mappings defined earlier 
are special cases of our general mapping. 
The correspondence of section IIC 
is recovered by setting $i=\de=j=0$ in the Coulomb system
and $I=\De=J=0$ in the oscillator system,
so that $A=2a=0$.
Figure 2 is a commutative diagram 
showing the relationship between 
this simpler mapping and the general mapping.
Similarly,
the mapping of section IIF
between the fermionic sector 
of the supersymmetric Coulomb system
and the second-fermionic sector 
of the supersymmetric oscillator 
is reproduced with the choices 
$i=1$, $\de=0$, $j=0$ 
and $I=2$, $\De=0$, $J=0$,
so that $A=2a=2$.

For the case of constant 
nonnegative integral $A$ and $a$,
\eq{jje} controls the relationship 
between the supersymmetric sectors of the two systems.
While any iteration of the supersymmetry 
for the Coulomb system can be taken,
only {\em even} iterations
of the oscillator supersymmetry appear.
This is why we introduced the second-fermionic sector
of the supersymmetric oscillator in section IIF.
It is therefore possible to combine Figures 1 and 2
in a single commutative diagram.
Moreover,
the general mapping shows that
Figure 1 can be extended downward
to incorporate higher iterations of the supersymmetry.
The result is an infinite series of mappings relating
Coulomb systems with $a=2$, $3$, $4$, $\ldots $
to oscillator systems with $A=4$, $6$, $8$, $\ldots$, 
respectively.

\vglue 0.6cm
{\bf\ni IVB. 
Three-Dimensional Coulomb and Oscillator Systems}
\vglue 0.4cm

To obtain further insight
about the content of the general mapping of section IVA,
we next restrict attention to the special case 
where both systems are three-dimensional.
Since this choice can be implemented with $j=0$,
we assume this in what follows.

The general mapping becomes 
\bea
 W_{3,N^*,L^*}(Y) &=&
     K_{3,n^*,\alf} 
     Y^{-\alf} 
     w_{3,n^*,l^*} \left(n^*Y^2\right)         
                                    \quad ,\label{33func} \\
N^* &=& 2 n^*- \frac 3 2            \quad ,\label{33n}    \\
L^* &=& 2 l^*+\half                 \quad , \label{33l}   \\
\De-I &=& 2(\de-i)+\la-\half      \quad . \label{33del}
\eea
The values of $\la$ allowed by \eq{jjc} are $\la=0, 1$. 
The orthonormality requirements
\eq{cRange} and \eq{oRange} 
become
\bea
\de -i &<& l+1          \quad , \label{d3cond} \\
\De-I &<& L+ \frac 3 2 \quad . \label{D3cond}
\eea
We can regard \eq{d3cond} and \eq{D3cond} as
conditions limiting the choice of quantum defects 
in the two systems to 
a semi-infinite region of the $(\De-I)$ versus $(\de-i)$ plane. 
The condition \rf{33del}
then further restricts the choice 
to a straight line in this region. 

One interesting special case 
is obtained by requiring that the oscillator be exact 
in the sense that $\De-I=0$.
Then,
\eq{33del} becomes 
\beq
\de-i \ = \ -a\ = \ \half(\half-\la) \quad ,
\eeq
showing that a nonzero defect 
in the Coulomb system is necessary.
The eigenvalues of the equations are 
\bea
E_{n^*,\ga^*} &=& \fr{-E_0}{(2n+\la-\half)^2} \quad , \\
F_{N,\Ga}       &=&  F_0 (2N+2\la + 2) 
\quad ,
\eea
and the relationships among 
the principal quantum numbers 
and the angular momenta become
\bea
L &=& 2l^* + \half \quad , \label{33cl} \\
N &=& 2n^* - \frac 3 2 \quad .
\eea
Selecting $\la=1$ for definiteness,
we see that \eq{33cl} maps each 
successive Coulomb angular momentum
$l^*=\frac 1 4$, $l^*=\frac 5 4$, $l^*=\frac 9 4$, $\ldots $
to every second oscillator angular momentum
starting at $L=\la$:
$L = 1$, $L=3$, $L=5$, $\ldots$.
The mapping therefore preserves the degeneracy of states.
For instance, the kets 
$\ket{n^*=\frac 9 4, \ l^*=\frac 1 4}$ and
$\ket{n^*=\frac 9 4, \ l^*=\frac 5 4}$, 
which are degenerate in the Coulomb system,
are mapped to the degenerate states
$\ket{N=3, \ L=1}$ and  
$\ket{N=3, \ L=3}$ 
in the oscillator system.
This feature is also a characteristic of 
the original mapping of section IIC.
The main differences here are that 
the Coulomb effective angular momenta $l^*$
are nonintegral and, more importantly,
that both systems are three dimensional.

A second case of interest is obtained when
the Coulomb system is exact, 
i.e.,
$\de-i=0$. 
The condition $j=0$ implies that $a=0$ too.
Then,
\eq{jje} becomes $\De-I = \la - \half$,
showing that a nonzero defect is again needed,
this time in the oscillator system.
The eigenvalues are 
\bea
E_{n,\ga}       &=& \fr{-E_0}{4n^2}            \quad , \\
F_{N^*,\Ga^*} &=&  F_0 (2N-2\la+4)             \quad ,
\eea
and the mapping gives 
\bea
L^* &=& 2l + \half \quad , \label{33ol} \\
N^* &=& 2n - \frac 3 2 \quad .
\eea
The first of these equations
shows every second oscillator angular momentum is imaged,
which again preserves the degeneracy of states.
If,
for example,
we choose $\la=0$, 
then the degenerate Coulomb states
$\ket{n=3, \, l=1}$ and $\ket{n=3, \, l=2}$
map to the degenerate oscillator states 
$\ket{N^*=\frac 9 2, \, L^*=\frac 5 2}$
and 
$\ket{N^*=\frac 9 2, \, L^*=\frac 9 2}$.

In the above examples,
the quantities $\De-I$ and $\de-i$
are constant.
In physical systems such as alkali-metal atoms,
$\de-i$ depends on $l$ and 
tends towards zero as $l$ increases.
This feature can also be incorporated in
our general mapping.
It implies a dependence of $\De-I$ on $L$,
which might reflect a realistic feature of
a physical oscillator
such as a cloud of particles caught in a Penning trap.

As an example, 
we map the SQDT radial equation for the physical sodium atom
into a three-dimensional SQDT oscillator.
In sodium,
the inaccessibility of the levels occupied by the 
ten inner electrons is implemented by the choices
$i(0)=2$, $i(1)=1$, and $i(l\geq 2)=0$.
The quantum defects $\de(l)$ in this case
are known 
\cite{hk}.
Choosing for definiteness $\la=1$ and
selecting $I(0)=2$, $I(1)=1$, $I(L\geq 2)=0$,
the values of $\De$ can be found from \eq{33del}.
Table 1 lists the results.
As expected,
the values of $\De$ tend towards $\half$ as $L$ increases.

\vglue 0.6cm
{\bf\ni V. Mappings for Continuum States}
\vglue 0.4cm

In previous sections,
we have explored mappings between 
the bound states of the Coulomb and oscillator systems. 
It is natural to consider whether similar mappings exist
taking the unbound Coulomb states 
into an appropriate oscillator.
This question is of lesser physical interest at present,
so we restrict ourselves here to a brief 
outline of a possible approach to this issue.

The Coulomb problem with energies $E>0$
can be viewed as a scattering problem.
Following the general procedure of section IIA
again yields the differential equation
\rf{a5}, 
but with $E>0$.
In terms of the confluent hypergeometric function ${_1}F{_1}$,
the solutions are
\bea
w_{d,E,l}(y) &\propto &y^{l+\ga+1} 
\exp\left(\pm i y\sqrt{\fr E{E_0}} \right) 
\nonumber\\
&& \times  ~ 
{_1}F{_1}\left(l+\ga+1 \pm \fr 1 {2 i} \sqrt{\fr{E_0}E}, 
2(l+\ga+1),\mp 2 i y\sqrt{\fr E{E_0}} \right)
\quad .
\label{k6}
\eea
The upper and lower signs 
correspond to outgoing and incoming waves,
respectively.
The results of section IIA can be recovered
by taking $E$ to be negative
and choosing the upper sign.

It turns out that 
the appropriate image oscillator system 
\cite{bsw}
is the {\em inverted} oscillator,
with potential 
$U(R) = - \half M \Om R^2$.
This system is unbound.
The procedure of section IIB
gives a differential equation 
identical to \eq{b3} except that 
the sign of the potential $Y^2$ is reversed.
The solutions $W_{D,F,L}(Y)$ are 
\beq
W_{D,F,L}(Y) \propto Y^{L+\Ga+1} \, 
\exp\left(\pm \half i Y^2\right) \,
{_1}F{_1}\left(\half(L+\Ga+\frac 3 2) \mp\frac{i F}{4F_0},
         L+\Ga+\frac 3 2 ,\mp i Y^2  \right)
\quad .
\label{l3}
\eeq
These functions may not be physically permissible,
but are relevant for the purposes of establishing a mapping.
With the choice of the upper sign,
the wave functions for the usual oscillator 
may be obtained up to a constant by the analytic continuation 
$Y^2 \rightarrow iY^2$ and 
$F \rightarrow -iF$. 

A correspondence analogous to the mapping of section IIC 
exists between the continuum Coulomb states 
and the inverted-oscillator functions.
It is
\bea
W_{D,F,L}(Y) &\propto & Y^{- \half} \  
w_{d,E,l} \left(\fr {Y^2} 
{2 \sqrt{E/E_0}}\right) \quad ,\label{m4} \\
Y^2 &=& 2y\sqrt{E/E_0} \quad ,\label{m1} \\
D &=& 2d - 2 - 2\la \quad , \\
L &=& 2 l + \la \quad ,\label{m2} \\
\fr F {F_0} &=& 2 \sqrt{\fr{E_0}{E}} \label{m3}
\quad .
\eea
There are many similarities between this mapping
and the one discussed in section IIC.
Again, $\la$ must be integral,
so only even-dimensional oscillators are available as images.
Also,
the angular momenta $L$ 
are restricted to being either all odd or all even, 
thus eliminating half the oscillator states.
However,
the energy relation \rf{m3}, unlike \rf{c4},
involves continuous values of $E$ and $F$.
It also reveals that negative energies $F$ 
are excluded from the mapping.
As one energy tends to zero the other 
tends to infinity.

As an aside,
we remark that the negative energies $F$ 
do appear when considering the {\em repulsive} Coulomb problem. 
The differential equation of this problem is mapped into 
the inverted-oscillator differential equation 
\cite{pc} by a map
with \rf{m4} through \rf{m2} 
unchanged but with a negative sign taken for the square root
in \eq{m3}.

Although this lies outside the scope of the present work,
it seems feasible that supersymmetry 
could be introduced into these systems
along with the corresponding SQDT.
We conjecture that this allows for odd dimensions $D$.
Since parabolic coordinates have some 
advantages for scattering problems,
it would also be interesting to perform
an analysis in terms of the dual 
parabolic-coordinate supersymmetries
of ref.\ \cite{bk}
instead of the spherical-coordinate supersymmetry used here.

\vfill\eject
{\bf\ni VI. Summary}
\vglue 0.4cm

In this paper,
we generalized the radial mappings
first identified by Schr\"odinger
that relate the Coulomb and oscillator systems.
Our principal result is
a mapping between the supersymmetry-based quantum-defect
theories for the Coulomb and oscillator systems
in arbitrary dimensions.
In particular,
odd oscillator dimensions can be accessed
as well as the usual even ones.
The mapping and some of its 
limits are illustrated in Figures 1 and 2.

In deriving this result,
we have extended to arbitrary dimensions 
the analytical SQDT in three dimensions
used to describe physical alkali-metal atoms.
An analogous SQDT for the harmonic oscillator
in arbitrary dimensions has also been presented.
In suitable limits,
these theories reproduce the bosonic and fermionic 
sectors of the corresponding 
supersymmetric quantum-mechanical systems.
We have elucidated a basic relationship between
the supersymmetric radial Coulomb and oscillator systems:
the $q$th iteration of supersymmetry for the Coulomb system
corresponds naturally to
the $2q$th iteration of supersymmetry for the oscillator.
For the special case of the one-dimensional radial oscillator,
we uncovered a quantum-mechanical supersymmetry 
in which the parity is restricted to be either odd or even.
We have also briefly considered
mappings relating the continuum-spectrum states of the 
Coulomb and oscillator systems.

The issue of the physical relevance of our results
has also been addressed in part.
The three-dimensional Coulomb SQDT is known to provide 
a good analytical description 
of the behaviour of Rydberg atoms.
Our mapping provides a means of obtaining 
an equivalent analytical oscillator SQDT.
An explicit example mapping the sodium atom
to an oscillator SQDT is given in Table 1.
It is also possible that 
an oscillator SQDT could be used to describe
a suitable physical system,
perhaps the Penning trap.
If this can be realized in practice,
the generalized mapping presented here
could provide a connection between two
apparently disparate physical systems.

\vglue 0.6cm
{\bf\ni VII. Acknowledgments}
\vglue 0.4cm

We thank Larry Biedenharn for discussion.
This work was supported in part 
by the United States Department of Energy
under grant number DE-FG02-91ER40661. 

\vglue 0.6cm
{\bf\ni VIII. References}
\vglue 0.4cm

\def\ajp #1 #2 #3 {Am.\ J.\ Phys.\ {\bf #1}, #3 (19#2)}
\def\ant #1 #2 #3 {At. Dat. Nucl. Dat. Tables {\bf #1}, #3 (19#2)}
\def\ap #1 #2 #3 {Ann.\ Physics\ {\bf #1}, #3 (19#2)}  
\def\ijqc #1 #2 #3 {Internat.\ J.\ Quantum\ Chem.\  
  {\bf #1}, #3 (19#2)}   
\def\jmp #1 #2 #3 {J.\ Math.\ Phys.\ {\bf #1}, #3 (19#2)}
\def\jms #1 #2 #3 {J.\ Mol.\ Spectr.\ {\bf #1}, #3 (19#2)}  
\def\jpa #1 #2 #3 {J.\ Phys.\ A {\bf #1}, #3 (19#2)}   
\def\jpsj #1 #2 #3 {J.\ Phys.\ Soc.\ Japan {\bf #1}, #3 (19#2)}
\def\lnc #1 #2 #3 {Lett.\ Nuov.\ Cim. {\bf #1}, #3 (19#2)}
\def\mj #1 #2 #3 {Math. Japon. {\bf #1}, #3 (19#2)}   
\def\mpl #1 #2 #3 {Mod.\ Phys.\ Lett.\ A {\bf #1}, #3 (19#2)}
\def\nat #1 #2 #3 {Nature {\bf #1}, #3 (19#2)}
\def\nc #1 #2 #3 {Nuov.\ Cim.\ A{\bf #1}, #3 (19#2)}
\def\nim #1 #2 #3 {Nucl.\ Instr.\ Meth.\ B{\bf #1}, #3 (19#2)}
\def\npb #1 #2 #3 {Nucl.\ Phys.\ B{\bf #1}, #3 (19#2)}
\def\pha #1 #2 #3 {Physica \ {\bf #1}, #3 (19#2)}
\def\pjm #1 #2 #3 {Pacific J. Math. \ {\bf #1}, #3 (19#2)}   
\def\pla #1 #2 #3 {Phys.\ Lett.\ A {\bf #1}, #3 (19#2)}
\def\plb #1 #2 #3 {Phys.\ Lett.\ B {\bf #1}, #3 (19#2)}
\def\prep #1 #2 #3 {Phys.\ Rep. {\bf #1}, #3 (19#2)}  
\def\pra #1 #2 #3 {Phys.\ Rev.\ A {\bf #1}, #3 (19#2)}  
\def\prd #1 #2 #3 {Phys.\ Rev.\ D {\bf #1}, #3 (19#2)}
\def\prl #1 #2 #3 {Phys.\ Rev.\ Lett.\ {\bf #1}, #3 (19#2)}
\def\prs #1 #2 #3 {Proc.\ Roy.\ Soc.\ (Lon.) A {\bf #1}, #3 (19#2)}
\def\ptp #1 #2 #3 {Prog.\ Theor.\ Phys.\ {\bf #1}, #3 (19#2)}
\def\ibid #1 #2 #3 {\it ibid., \rm {\bf #1}, #3 (19#2)}

\newpage
\vspace*{1cm}\noindent
Table 1. Possible parameters for a mapping 
between the radial Coulomb and oscillator systems 
in three dimensions.
For the choices $j=0$ and $\la=1$,
values of $l$, $i$, $n$, $n_s$, and $\de$ 
are tabulated for sodium
along with the corresponding values 
of $L$, $I$, $N$, $N_s$, and $\De$ 
under the mapping \rf{33del}.
The quantities $I(L)$ have been selected
to fill all levels below $N_s=5$. 

\vspace*{1cm}\noindent 
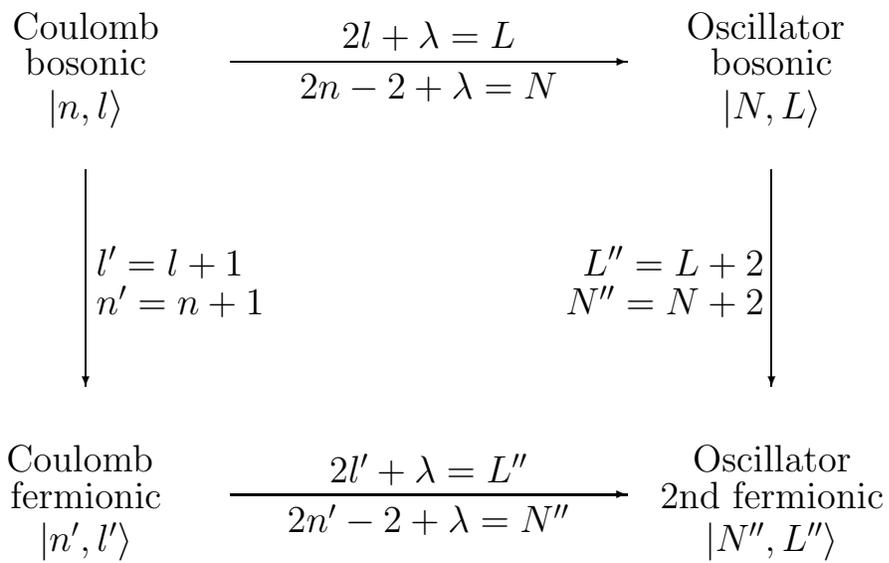
Figure 1. Supersymmetric mappings. 
Relationships are shown interconnecting 
the bosonic and fermionic
partners of the Coulomb and oscillator systems.
The diagram is commutative.

\vspace*{1cm}\noindent
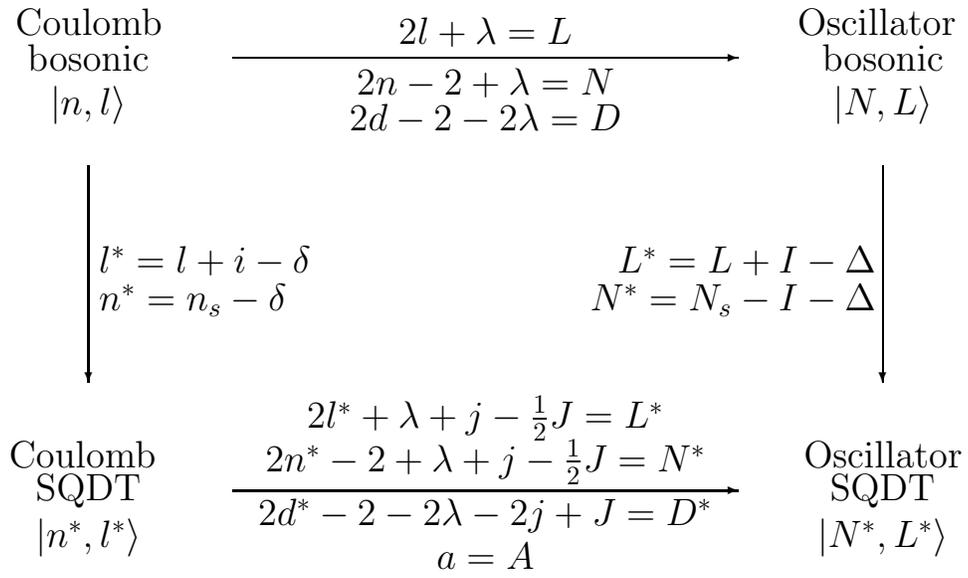
Figure 2. SQDT mappings. 
Relationships are shown interconnecting the bosonic 
and SQDT sectors of the Coulomb and oscillator systems.
The diagram is commutative.

\newpage
\center{Table 1}
\vspace*{3cm}
\begin{figure}
\large
\begin{displaymath}
\begin{array}{|r|c|l|l|l||r|c|c|c|l|}
\hline
\multicolumn{5}{|c||}{\mbox{Coulomb system (sodium)}} &
\multicolumn{5}{c|}{\mbox{Oscillator system}} \\
\hline
\multicolumn{1}{|c|}{l} & i & \multicolumn{1}{c|}{n} &  
  \multicolumn{1}{c|}{n_s} & \multicolumn{1}{c||}{\de} & 
  \multicolumn{1}{c|}{L} & I & N & N_s & 
  \multicolumn{1}{c|}{\De} \\
\hline
0      & 2 & \geq 1   & \geq 3 
& 1.35  & 1    & 2 & \geq 1 & \geq 5 & 1.20  \\
1      & 1 & \geq 2   & \geq 3 
& 0.859 & 3    & 1 & \geq 3 & \geq 5 & 1.218 \\
2      & 0 & \geq 3   & \geq 3 
& 0.01  & 5    & 0 & \geq 5 & \geq 5 & 0.52  \\
3      & 0 & \geq 4   & \geq 4 
& 0.00  & 7    & 0 & \geq 7 & \geq 7 & 0.50  \\
\geq 4 & 0 & \geq l+1 & \geq l+1 
& 0 & \geq 9 & 0 & \geq L & \geq L & 0.5 \\
\hline
\end{array}
\end{displaymath}
\normalsize
\end {figure}

\newpage
\center{Figure 1}
\vspace*{3cm}
\begin{figure} 
\setlength{\unitlength}{4.8mm}
\large
\begin{picture}(28,30)(-5,12)

\put(2,29){\makebox(0,0){Coulomb} }
\put(2,28){\makebox(0,0){bosonic}}
\put(2,26.7){\makebox(0,0){$\ket{n,l}$}}

\put(2,17){\makebox(0,0){Coulomb }}
\put(2,16){\makebox(0,0){fermionic}}
\put(2,14.7){\makebox(0,0){$\ket{\pr n,\pr l}$}}

\put(21,29){\makebox(0,0){Oscillator }}
\put(21,28){\makebox(0,0){bosonic}}
\put(21,26.7){\makebox(0,0){$\ket{N, L}$}}

\put(21,17){\makebox(0,0){Oscillator}}
\put(21,16){\makebox(0,0){2nd fermionic}}
\put(21,14.7){\makebox(0,0){$\ket{\ppr N,\ppr L}$}}

\put(6,28){\vector(1,0){11}}  
\put(6,16){\vector(1,0){11}}   

\put(2,25){\vector(0,-1){6}}
\put(21,25){\vector(0,-1){6}}   

\put(9,28.2){\makebox(5,1){$2l+\la = L$}}
\put(9,26.8){\makebox(5,1){$2n-2+\la = N$}}
\put(9,16.2){\makebox(5,1){$2\pr l+\la = \ppr L$}}
\put(9,14.8){\makebox(5,1){$2\pr n -2 +\la = \ppr N $}}
\put(2.3,21){\shortstack[l]{$\pr l = l+1$ \\ $\pr n = n+1$}}
\put(15.3,21){\shortstack[r]{$\ppr L = L+2$ \\ $\ppr N = N+2$}}
\end{picture}
\normalsize
\setlength{\unitlength}{1pt}
\end{figure}


\newpage
\center{Figure 2}
\vspace*{3cm}
\begin{figure} 
\setlength{\unitlength}{4.8mm}
\large
\begin{picture}(28,30)(-3,12)

\put(2,29){\makebox(0,0){Coulomb} }
\put(2,28){\makebox(0,0){bosonic}}
\put(2,26.7){\makebox(0,0){$\ket{n,l}$}}

\put(2,17){\makebox(0,0){Coulomb }}
\put(2,16){\makebox(0,0){SQDT}}
\put(2,14.7){\makebox(0,0){$\ket{ n^*,l^*}$}}

\put(24,29){\makebox(0,0){Oscillator }}
\put(24,28){\makebox(0,0){bosonic}}
\put(24,26.7){\makebox(0,0){$\ket{N, L}$}}

\put(24,17){\makebox(0,0){Oscillator}}
\put(24,16){\makebox(0,0){SQDT}}
\put(24,14.7){\makebox(0,0){$\ket{N^*,L^*}$}}

\put(6,28){\vector(1,0){14}}  
\put(6,16){\vector(1,0){14}}   

\put(2,25){\vector(0,-1){6}}
\put(24,25){\vector(0,-1){6}}   

\put(10.5,28.2){\makebox(5,1){$2l+\la = L$}}
\put(10.5,26.8){\makebox(5,1){$2n-2+\la = N$}}
\put(10.5,25.8){\makebox(5,1){$2d-2-2\la = D$}}

\put(10.5,17.5){\makebox(5,1){$2l^*+\la+j-\half J = L^*$}}
\put(10.5,16.3){\makebox(5,1){$2n^*-2+\la+j-\half J = N^* $}}
\put(10.5,14.8){\makebox(5,1){$2d^*-2-2\la-2j+J = D^*$}}
\put(10.5,13.7){\makebox(5,1){$a = A$}}
\put(2.3,21){\shortstack[l]{$l^* = l+i-\de$ \\ $n^* = n_s-\de$}}
\put(15.9,21){\shortstack[r]{$L^* = L+I-\De$ \\ $N^* = N_s-I-\De$}}
\end{picture}
\setlength{\unitlength}{1pt}
\end{figure}



\begin{thebibliography}{xx}

\bibitem{lc1906}
T. Levi-Civita, 
Act. Math. {\bf 30}, 305 (1906)

\bibitem{sch}
E. Schr\"odinger,
Proc. Roy. Irish Acad. Sect. A {\bf 46}, 183 \ (1941)

\bibitem{lo}
J.D. Louck,
\jms 4 60 334

\bibitem{bf}
D. Bergmann and Y. Frishman,
\jmp 6 65 1855

\bibitem{dm}
V.A. Dulock and H.V. McIntosh,
Pacific J. Math. {\bf 19}, 39 (1966)

\bibitem{rr}
R. Rockmore,
\ajp 43 75 29

\bibitem{clm}
E. Chac\'on, D. Levi and M. Moshinsky,
\jmp 17 76 1919

\bibitem{cp}
J. \v{C}\'{\i}\v{z}ek and J. Paldus,
\ijqc 12 77 875

\bibitem{knt}
V.A. Kosteleck\'y, M.M. Nieto and D.R.Truax,
\prd 32 85 2627

\bibitem{ak}
V.A. Kosteleck\'y,
p.\ 295 in {\it Symmetries in Science VII}, 
eds. B. Gruber and T. Otsuka
(Plenum, New York, 1994)
(quant-ph/9508015)

\bibitem{ni}
H. Nicolai,
\jpa 9 76 1497 

\bibitem{wi}
E. Witten,
\npb 185 81 513

\bibitem{kn1}
V.A. Kosteleck\'y and M.M. Nieto,
\prl 53 84 2285 ;
\pra 32 85 1293 .

\bibitem{kn}
V.A. Kosteleck\'y and M.M. Nieto,
\pra 32 85 3243 .
For a review and references to related literature,
see, for example,
ref.\ \cite{ak}.
Recent developments in the 
subject are discussed in
R. Bluhm and V.A. Kosteleck\'y,
\pra 49 94 4628 
(quant-ph/9508020);
\ibid 50 94 R4445 
(hep-ph/9410325);
\ibid 51 95 4767 
(quant-ph/9506009);
R. Bluhm, V.A. Kosteleck\'y and B. Tudose,
\pra 52 95 2234 
(quant-ph/9509010);
\it ibid., \rm in press
(quant-ph/9510023).

\bibitem{ks}
P. Kustaanheimo and E. Stiefel, \
J. Reine Angew. Math. \ {\bf 218}, \ 204 \ (1965)

\bibitem{cm}
A. Cisneros and H.V. McIntosh,
\jmp 10 69 277

\bibitem{bo73}
M. Boiteux,
\pha 65 73 381

\bibitem{ch80}
A.C. Chen,
\pra 22 80 333 ; 2901;
\jmp 23 82 412

\bibitem{kin83}
M. Kibler and T. Negadi,
\lnc 37 83 225

\bibitem{dmpst}
L.S. Davytan, L.G. Mardoyan, G.S. Pogosyan,
A.N. Sissakian and \\ V.M. Ter-Antonyan,
\jpa 20 87 6121

\bibitem{lk88}
D. Lambert and M. Kibler,
\jpa 21 88 307

\bibitem{hk}
H.G. Kuhn,
{\em Atomic Spectra}
(Academic, New York 1969)

\bibitem{bsw}
A.O. Barut, C.K.E. Schneider
and R. Wilson,
\jmp 20 79 2244

\bibitem{pc}
P. Collas,
\jmp 22 91 2512

\bibitem{bk}
R. Bluhm and V.A. Kosteleck\'y,
\pra 47 93 794

\end{thebibliography}
\end{document}